\documentclass[10pt,onecolumn,nofootinbib]{revtex4}
\usepackage{graphicx}
\usepackage{textcomp}
\usepackage{dcolumn} 
\begin{document}

\title{Cover slip external cavity diode laser}
\author{Adra V. Carr, Yancey H. Sechrest, Scott R. Waitukaitis, John D. Perreault$^a$, Vincent P.A. Lonij and Alexander D. Cronin}
\affiliation{University of Arizona, Department of Physics, Tucson,
AZ, 85721}
\affiliation{$^a$University of Colorado, JILA, Boulder, CO 80309-0440}

\date{July 26, 2007}

\begin{abstract}
The design of a 671 nm diode laser with a mode-hop-free tuning range of 40 GHz is
described. This long tuning range is achieved by simultaneously
ramping the external cavity length with the laser injection current. 
The external cavity consists of a microscope cover slip mounted on
piezoelectric actuators. In such a configuration the laser output pointing remains fixed, independent of its frequency. Using a diode with an output power of 5-7 mW, the laser linewidth was found to be smaller than 30 MHz. This cover slip cavity and feedforward laser current control system is
simple, economical, robust, and easy to use for spectroscopy, as we
demonstrate with lithium vapor and lithium atom beam experiments.

\end{abstract}

\maketitle{}

External cavities are often used to control the frequency of a
laser. Early investigations with a plane mirror reflecting light
back into a laser (the design in figure 1) showed that optical
feedback can ``pull" the laser frequency and influence which
laser-cavity mode is active \cite{SAL79,COK84}. Instead of a mirror a reflective grating can be used, as in the popular Littrow
\cite{MSW92,RWE95,CCD99,ALW00,PAW91,BBH91,WIH91} or Littman-Metcalf
\cite{STS06,HAM91,LIM78,WIH91} configurations with diode lasers.
However, scanning the cavity alone offers a limited mode-hop-free tuning
range with red diode lasers. This was reported by
\cite{WIH91,CCD99,BBH91}, and confirmed in our laboratory. We found
that 2 GHz was the maximum single-mode tuning range we could achieve
when using only optical feedback to control the frequency of a 671
nm diode laser. We tried seven different diode laser
models \cite{laser}, and found the same
limited tuning range with either a grating or a mirror forming the
external cavity. Adjusting only the diode laser current also
resulted in a limited (2-3 GHz) mode-hop-free tuning range. This scan range is
limited by the overlap of the modes for the laser diode and external
cavity. A substantial increase in the tuning range was caused by modulating
\emph{both} the external cavity and the diode laser current
together.

\begin{figure}[h]
\centering
\includegraphics[width=4.5in]{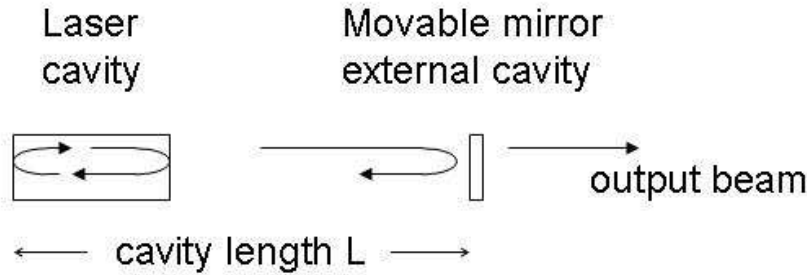}
\caption{The external cavity.  A movable mirror reflects some light
back to the laser. This builds up the power for circulating light
with a wavelength that is resonant with both the laser cavity and
the external cavity.}\label{fig:p2}
\end{figure}

Here we describe a diode laser system in which the laser current and
the external cavity length are simultaneously ramped to achieve a 40
GHz mode-hop-free tuning range operating near the lithium atom 2S-2P
transition wavelength of 671 nm. A microscope cover slip mounted on a piezoelectric transducer serves as the external cavity and output
coupler, hence we refer to this as a \emph{cover slip cavity}
\cite{JON07}. Surprisingly, the modest reflectivity of the glass
cover slip surface with no coating (R $\approx4$\%) is sufficient to
force single-mode operation while the laser frequency scans
continuously over 40 GHz. Since either end of this mode-hop-free tuning range
can be extended by adding (or subtracting) an offset to the cavity length, this mode-hop-free tuning range appears to be limited by non-linearity in the required diode laser current or cavity length adjustments.

It was found empirically that the laser current affects the
frequency with the slope \mbox{$dI/d\nu =$ -250 $\mu$A/GHz} for the
Sanyo model DL3149-057 diode laser.  The cavity length affects the
frequency with the slope \mbox{$dL/d\nu = $ -22 nm/GHz}.  This is
consistent with the relationship $ \Delta L/L = - \Delta \nu / \nu$
for our cavity (length $L=$10 mm) and red light ($\nu$=447 THz).
Therefore, as the cavity length increases by 22 nm, the laser current
should be increased by 250 $\mu$A to anticipate the -1 GHz shift in
laser frequency. This ``feedforward'' procedure increases the tuning
range of the cover slip cavity laser. As a feedforward system, there
is no feedback loop connecting the laser current and cavity length
adjustments; they are just commensurately scanned in a predictable
fashion using analog electronics. The temperature of the laser
mounting block was also stabilized to within 0.1 \textcelsius~
because temperature was seen to affect the laser frequency with the
slope $dT/d\nu$ = -0.03\textcelsius/GHz (i.e. a wavelength shift of 1
nm per 20\textcelsius ).

The laser current is modulated by an input voltage $V_{in}$ that can
be provided by a function generator.  A resistor in series with the
laser keeps the current linearly related to the input voltage [$I =
(V_{in} - V_{diode})/R$ to first order above the diode knee].  The same input
voltage is also used to control the cavity length. To accomplish
this, $V_{in}$ is added to an offset ($V_{offset}$), amplified with
a gain factor $G$, and then directly applied to the piezoelectric
actuators to adjust the cover slip cavity length. Hence the cavity
length [$L = L_0 + (V_{in}+V_{offset})G (dL/dV)$] is also linearly
related to the input voltage. The length-to-voltage multiplier
($dL/dV$=4 $\mu$m/150 V) of the piezoelectric
actuators \cite{piezo} makes the required
piezo-voltage to laser current slope -3.3V/mA for mode-hop-free laser
scanning as illustrated in Figure 2.

\begin{figure}
\centering
\includegraphics[width=4.25in]{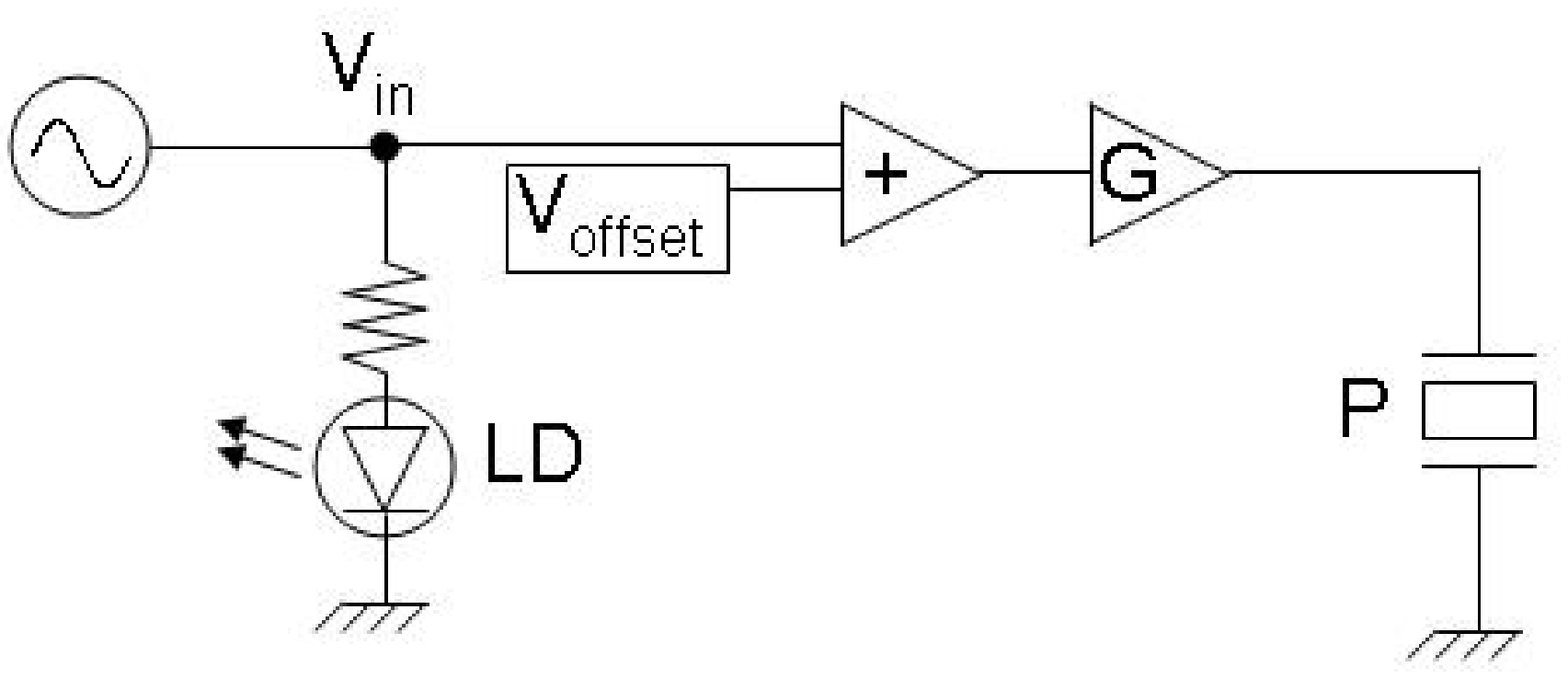}
\includegraphics[width=2.25in]{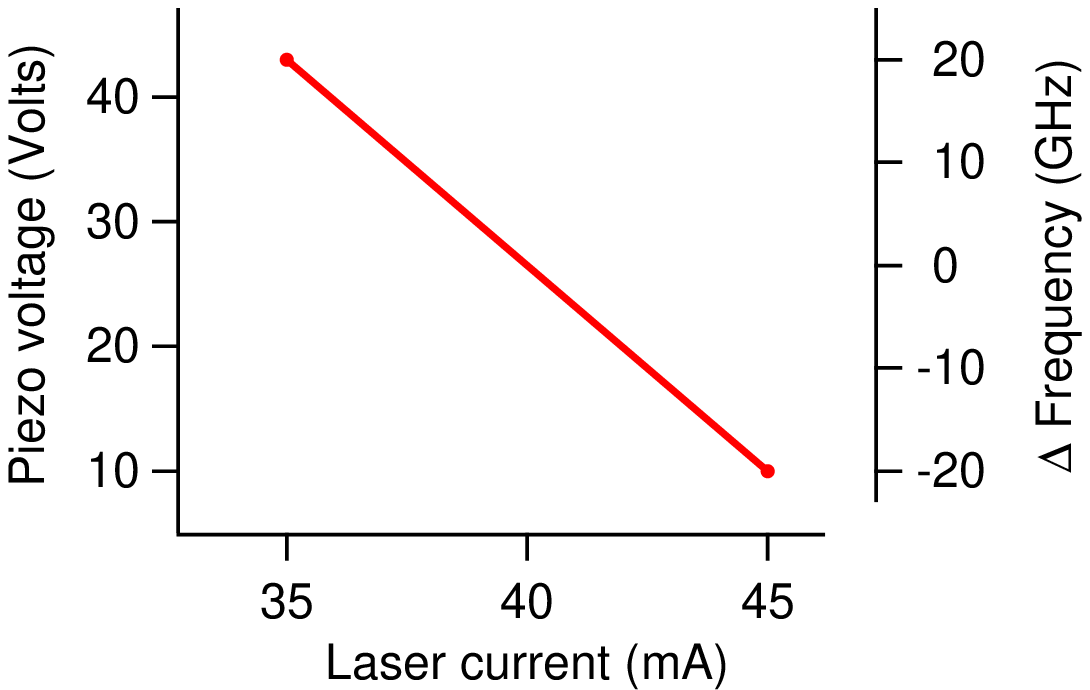}
\caption{(Left)  Feedforward electronics.  Current through the laser diode
(LD) depends on the input voltage ($V_{in}$). The same voltage is
shifted (by adding $V_{offset}$) then amplified by gain (G) and
applied to the 
piezoelectric actuator (P). Not shown: laser
protection circuitry. (Right) The predicted linear relation between the laser current and the cavity actuator voltage and the resulting shift in laser frequency. (The laser frequency is centered on 447,000 GHz for the wavelength 671
nm). Nonlinearities in the laser current as a function of
$V_{in}$, nonlinearities in the piezo-voltage as a function of $V_{in}$ or nonlinearities in the relationship of these quantities to laser frequency limit the mode-hop-free scan range of the system.}\label{fig2}
\end{figure}

Ramping the laser current simultaneously with the cavity length is
not a novel idea. This was also used to enhance the tuning range for the various
diode laser systems which utilized an external cavity grating \cite{ALW00,CCD99,RWE95,MSW92}. However, our cover slip
cavity with feedforward current control has the advantages of
simplicity, economy, and stability. We discuss these advantages
after explaining the mechanical construction.

\begin{figure}
\centering
\includegraphics[width=3.5in]{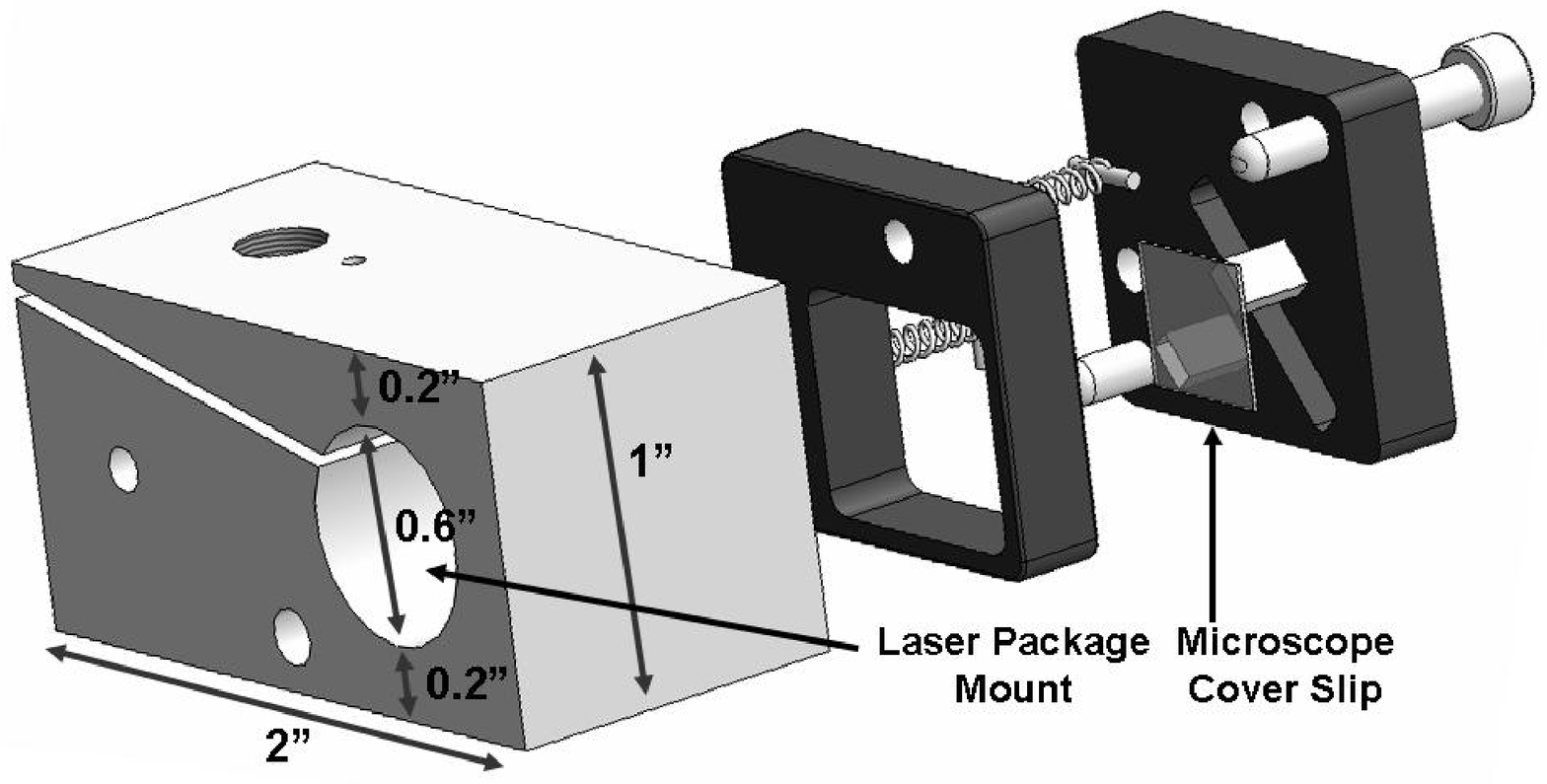}
\includegraphics[width=3.5in]{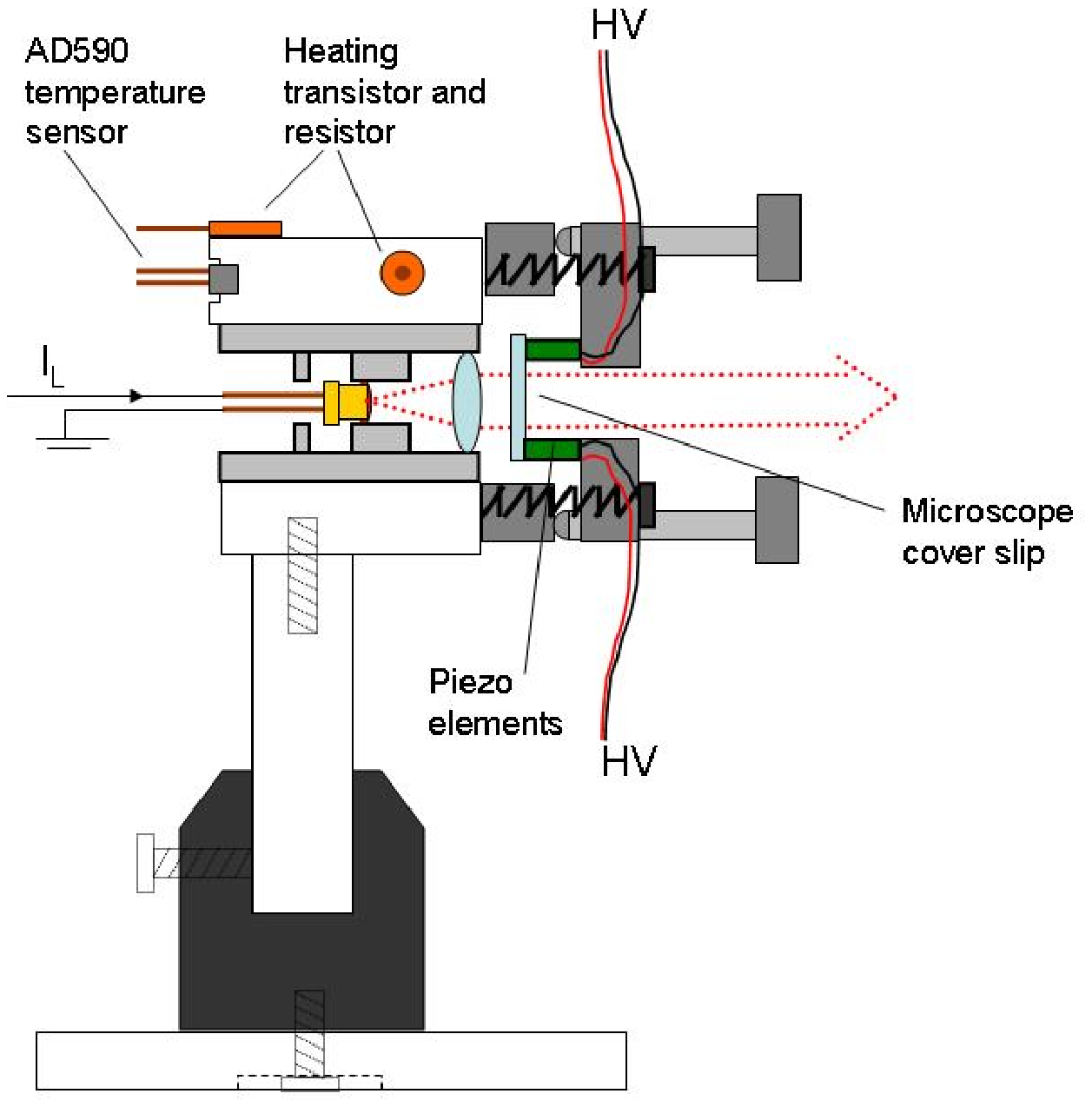}
\caption{(Left) Exploded view of the laser mounting block and cover
slip cavity assembly.  (Right) Side view of the laser
assembly.}\label{layout}
\end{figure}

Mechanical views of the cover slip cavity, collimating optics, and
temperature-stabilized laser mount are shown in Figure 3. The
physical layout has two principle parts: a mounting block, and a
kinematic mount. The laser diode is mounted in a
0.6\textsf{"}-diameter tube with collimating optics \cite{optics} and the tube is clamped in a
2\textsf{"}x1\textsf{"}x1\textsf{"} aluminum block pictured in
Figure \ref{layout}.  A temperature sensor (AD590) and heaters (a 5
Ohm resistor and 2 transistors) are also mounted on the block.

The stock 1\textsf{"}-square kinematic mount \cite{mount}
is disassembled and a 0.7\textsf{"}-square hole is milled in one
plate (as shown in Figure 3). A 0.3\textsf{"}-diameter hole is
drilled in the other plate to transmit the output beam. Two
piezoelectric actuators are mounted on either side of this output
hole, and a 0.5\textsf{"} diameter glass cover slip is attached to
the actuators with epoxy.  The cover slip is thus positioned within the kinematic mount so that it can form a short external cavity.  The kinematic mount is attached to the aluminum block with epoxy so that the cover slip and output hole are centered on the laser beam.  The mount enables coarse alignment of
the cover slip after the initial construction.
During feedforward operation, the system's only moving parts are the cover slip and the piezoelectric actuators.  The cavity
length can be coarsely adjusted (from 10 mm to 20 mm) by relocating
the collimating tube, and with care the kinematic mount can be taken
apart (to replace the cover slip) and reassembled while the
plate with the 0.7\textsf{"} hole remains attached to the mounting
block.

The block and cavity assembly is mounted on a standard optics post
and positioned in a thermally stable enclosure (with a hole for the
output beam). To reduce vibrations, the entire system is mounted on
a 1 Kg brass plate and placed on top of sorbathane pads.

One advantage of the cover-slip cavity is simplicity. If a grating
were used for the external cavity there are two mechanical design
criteria that must be considered because both the cavity length as
well as the grating angle affect the laser frequency. In comparison,
our cover slip cavity has only one important dimension: cavity
length. The laser beam always points in the same direction with the
cover slip cavity (whereas some grating cavities shift the beam
direction as the frequency is scanned). The cover slip transmits
$\sim 92\%$ of the laser beam so it offers a high output efficiency.  We
avoided anti-reflection coating the diode (as was done in
\cite{LBW95,CCD99,BBH91}), we avoided removing the diode can (which
has glass surfaces) and we avoided coating either side of the cover
slip (either to increase or decrease reflectivity) in order to keep
the design simple. The entire laser system was constructed from
parts that cost \$1500, and a lithium heat-pipe vapor cell was
constructed for a cost of \$800. The parts are listed in Appendix I.

If the temperature of the mounting block drifts by even
0.1\textcelsius~ then the offset voltage $V_{offset}$ needs to be
adjusted, but this is the only intervention needed to keep the
system scanning across the lithium atom resonance features for days
at a time.  This stability and long tuning range makes the red light
particularly useful for instructional laboratories and atomic
physics research. We operated individual Sanyo DL3149-057 lasers for
over 6 months with the current ramping in the range 35 to 45 mA. The
laser is typically scanned by applying a triangle waveform to
$V_{in}$ that varies from 8 to 10 volts and cycles at 100 Hz. For
some applications we also locked the laser frequency to a saturated
absorption feature by connecting $V_{in}$ to the output of a locking
circuit.

A number of experiments were done to test the performance of
the cover slip cavity system.  Lithium atom absorption spectra
demonstrating a 40-GHz mode-hop-free frequency scan are shown in
Figure 4. The vapor cell absorption features (Fig 4 upper left) each have a 3 GHz Doppler width
due to thermal motion of the lithium atoms. In this plot, data
are shown (lower left) for two vapor cell temperatures with correspondingly
different optical depths. The flat-bottomed absorption features with
the higher optical depth allowed us to measure the amount of
off-mode light in the laser output beam. This was seen to be 7 $\pm$
2 \%. The laser intensity is modulated from 5 to 7 mW in accord with
the laser current, but the laser beam pointing remains stable during
the scan. A Fabry-Perot transmission spectrum is shown in Figure 4 (lower left)
confirming the scanning range of 40 GHz. Saturated absorption
spectra shown in Figure 4 (upper right) were obtained using the two-probe method described in
\cite{ODL98,RRH98}. Also shown is an atom beam deflection signal (bottom right)
obtained by shining the cover slip cavity laser beam at a highly
collimated atom beam.  (See \cite{AA80} for a description of this
technique). Both types of Doppler-free spectra show the large 803
MHz hyperfine splitting due to the F=1,2 levels of the ground
$2S_{1/2}$ state of $^7$Li. The atom beam deflection spectrum also
shows a smaller but clearly resolved splitting due to the F=1,2
levels in the excited $2P_{1/2}$ state of $^7$Li.

\begin{figure}
\includegraphics[width=3.5in]{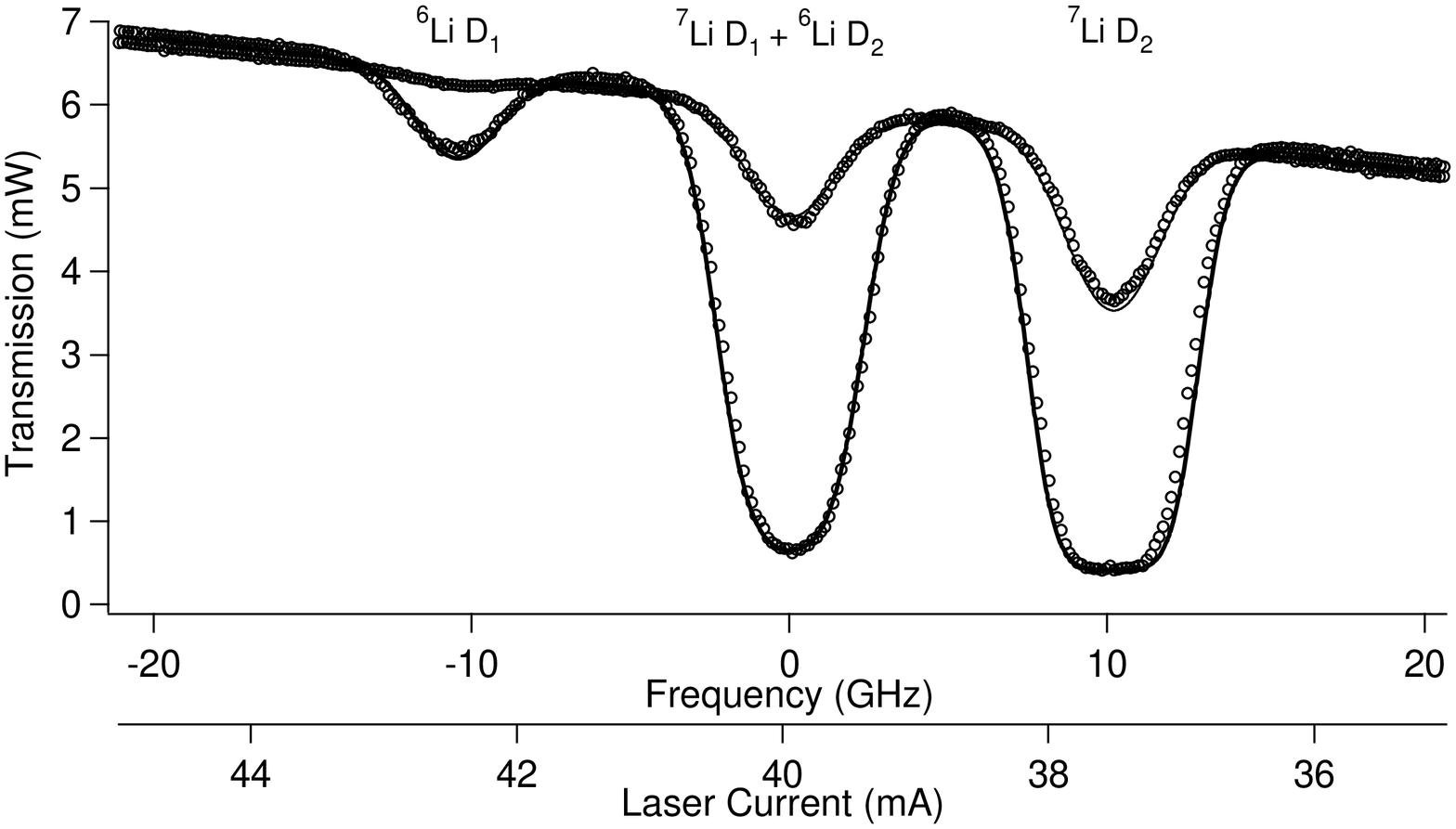}
\includegraphics[width=3.5in]{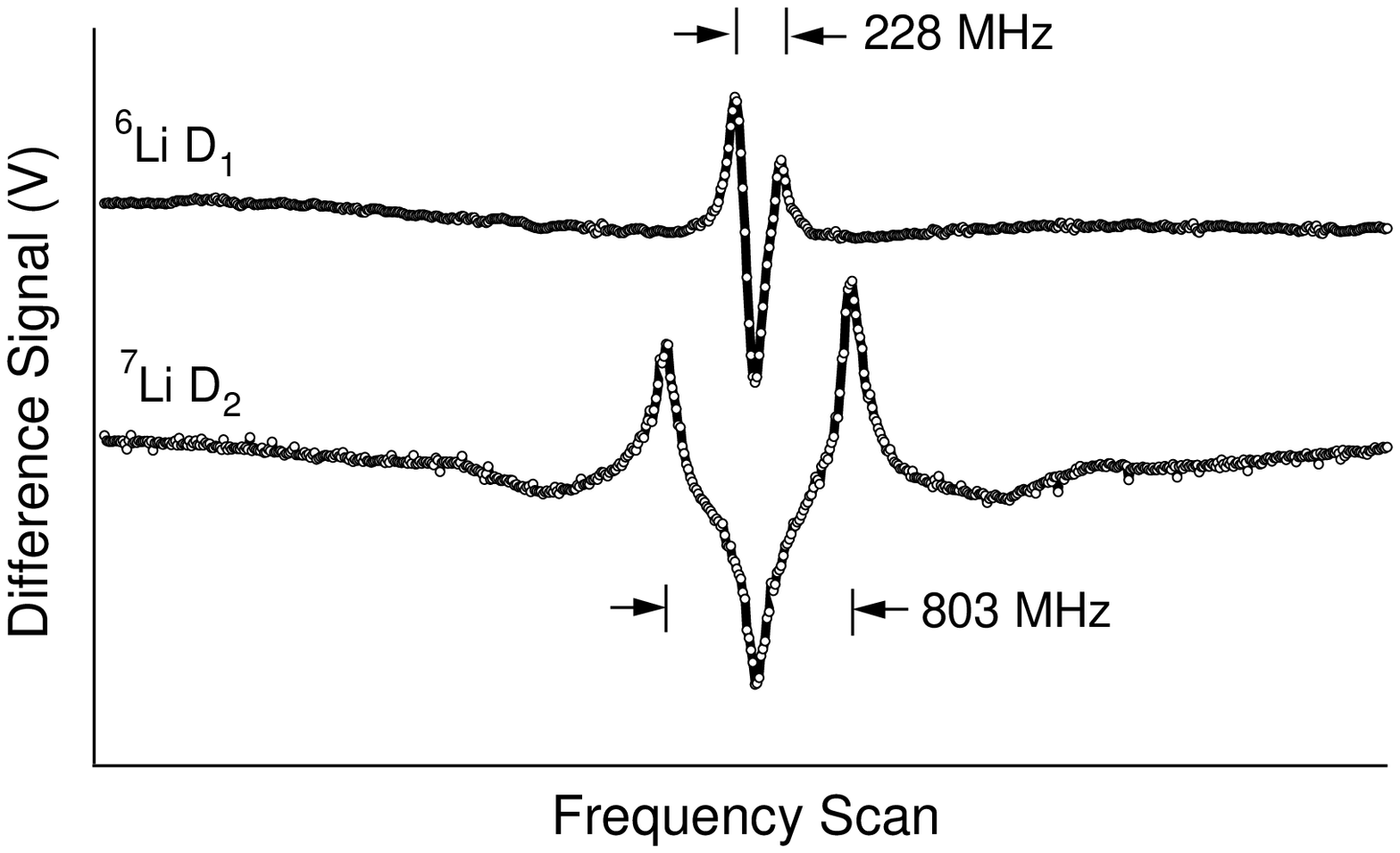}
\includegraphics[width=3.5in]{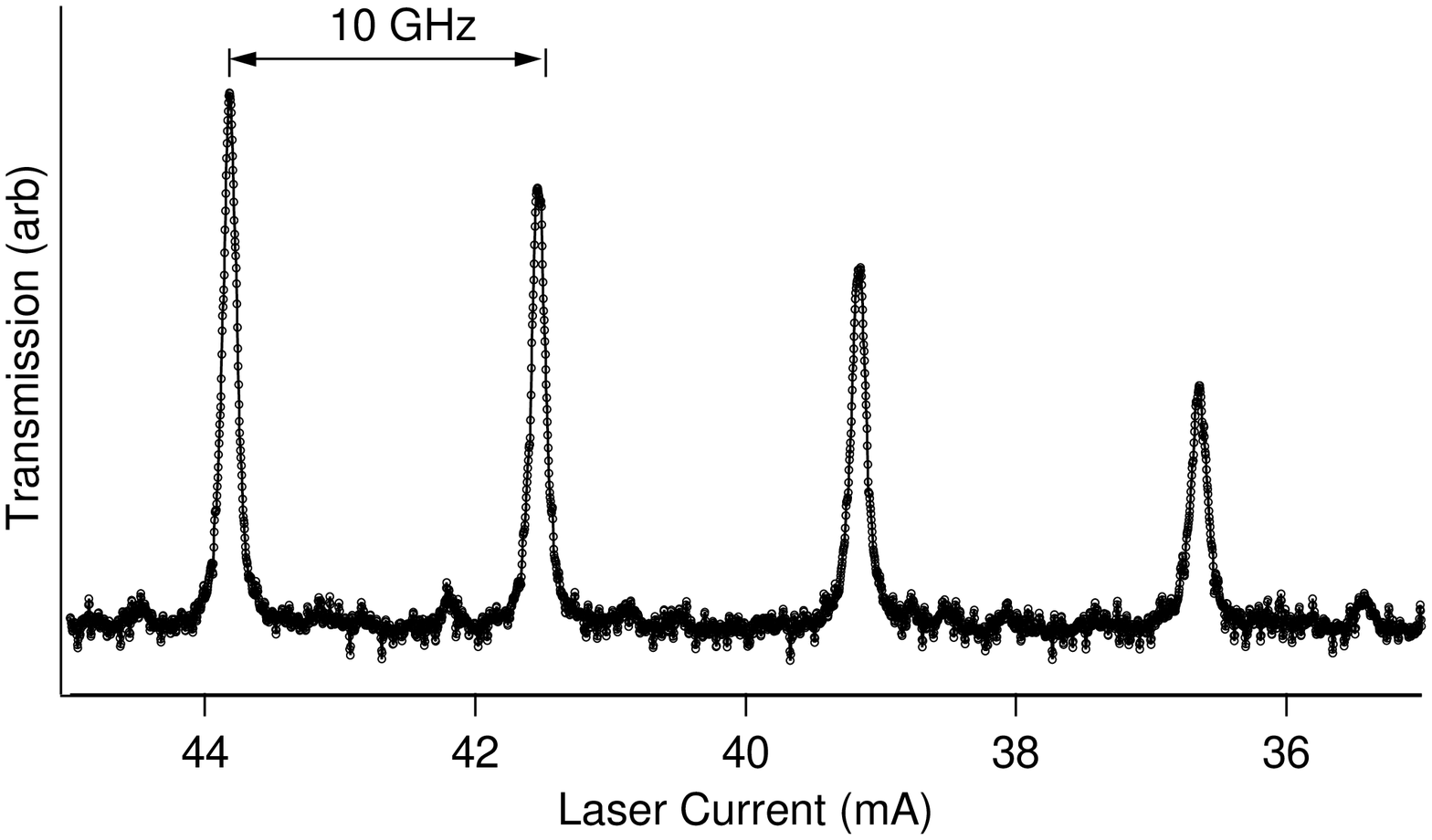}
\includegraphics[width=3.5in]{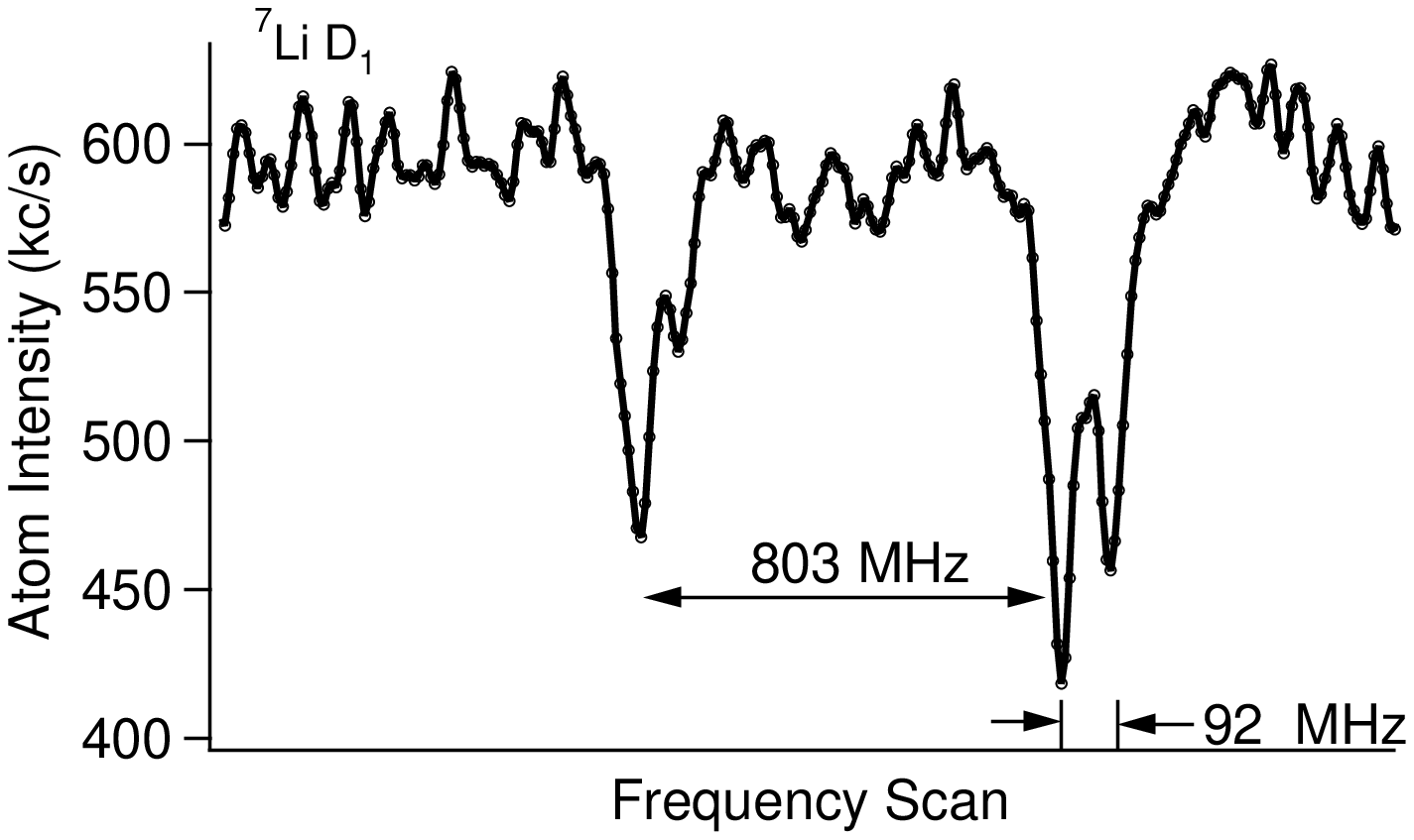}
\caption{(Top Left) Atomic transmission spectra
from lithium vapor. Two spectra (Top trace: T=370 \textcelsius, Bottom trace: T=420\textcelsius)
demonstrate that higher vapor temperature results in larger optical depth. (Bottom Left) Fabry-Perot transmission spectrum using a Fabry-Perot cavity. (Top Right) Saturated absorption signals.   The upward (increased transmission) peaks are
due to transitions out of the ground state hyperfine levels. (Bottom Right) Atom beam deflection spectrum.  Atom beam count rate is monitored as a function of laser frequency.  With the laser beam perpendicular to the atom beam, and the atom beam collimated to $10^{-4}$ radians, the
deflection spectrum has essentially zero Doppler width.  Power
broadening, transit time broadening, and the laser linewidth all
contribute to the four observed 55-MHz (FWHM) dips due to deflection. In addition to the 803 MHz spacing due to the ground state hyperfine splitting, the atom
beam deflection signal shows the 92 MHz hyperfine splitting due to the $2P_{1/2}$ excited state of $^7$Li.}\label{4fig}

\end{figure}

The laser line-width was smaller than 30 MHz, as found with a Fabry-Perot interferometer.  This upper limit on the linewidth is consistent with the atomic deflection spectra and the saturated absorption spectra shown in figure \ref{4fig}. In additional tests we modulated the laser current at 800 MHz (to add
sidebands to the laser spectrum) as described in \cite{WIH91}. This
did not reduce the tuning range.  We also used this laser to study
Faraday rotation in vapor \cite{VB96}, the Hanle effect in vapor \cite{SIL72}, and optical pumping of an atom beam.  We thus demonstrated that the cover slip
cavity diode laser serves for several atomic physics experiments.

In conclusion, we have described a diode laser system that uses a cover
slip cavity and feedforward operation to achieve a mode-hop-free
scanning range of 40 GHz.   The frequency can be scanned over the
lithium atom 2S-2P resonances near 671 nm, with the entire system being
simple, economical, stable and robust.

\section*{Appendix I: Parts and Supplies}
The following is a list of parts used to build the cover slip cavity
laser.
\begin{enumerate}
\item Sanyo Laser Diode: 670 nm, 7 mW, DL3149-057, \$14.14, and socket S7060,
\$3.98, Thorlabs
\item Collimation Tube with Lens, LT230P-B, \$108.00, and Spannor wrench, SPW301,
\$13.00, Thorlabs
\item Kinematic mount, KMS, \$33.50, Thorlabs
\item Piezoelectric elements (2), AE0203D04, \$72.00, Thorlabs
\item Piezoelectric controller, MDT694A,  \$680.00, Thorlabs
\item Function generator, GFG8020H, \$193.50, Mouser Electronics
\item Sorbathane Pad, SB12B, \$49.00, Thorlabs
\item Power supply (2), 680-WM113-D5, \$42.20, Mouser Electronics
\item Temperature sensor, AD590, \$12.80, Thorlabs
\item Thermocouple probe, TJC36CASS020K6SMPM, \$30.00, Omega Inc.
\item Thermocouple controller, HH74K, \$55.00, Omega Inc.
\end{enumerate}
Additional stock electronics components that were used to construct
the laser (op amps, breadboards, bud boxes, potentiometers), the
microscope cover slip, and laboratory supplies such as expoxy all
totaling \$120, are not itemized.
\\

The lithium heat-pipe oven was constructed from the following parts.
\begin{enumerate}
\item Optical viewports (2), FVG0150, \$95.26, Varian Inc.
\item 1.5" x 16" Stainless Steel tube w/ CF flanges and Nupro Inc. SS 4BG TSW valve, \$180.00, custom
welded
\item Heater wire, Aerocoax 1HN080B-3.9, \$3.50/ft, ARI Industries
Inc.
\item High temp cement, OB600, \$18.00, OMEGA Engineering INC.
\item  Variac voltage supply TDGC2-0.5, \$50.00, Circuit Specialists Inc.
\item Fibrafax thermal insulation, 93435K26, \$24.75/ft, McMaster-Carr
\item Thermocouple probe, TJC36CASS020K6SMPM, \$30.00, OMEGA
Engineering INC.
\item Thermocouple controller, HH74K, \$55.00, Omega Inc.
\item Li pellets (50g), Alfa Aesar \#41829, \$81.70, Sigma-Aldrich
Chemical

\end{enumerate}
The Li heat pipe oven operates with the central region near
400\textcelsius~and tube ends (and viewports) at 80\textcelsius.
Heater wire was wound around the central 5\textsf{"} of the 16\textsf{"} long Stainless Steel tube.
The central 8\textsf{"} were then covered in 2\textsf{"} thick fibrafax thermal
insulation.  The high temperature cement was used to bond the heater
wire to standard hookup wire outside the thermal insulation
surrounding the oven. Mounts for the heat pipe oven were machined
out of stock aluminum. Five grams of Li were loaded in the SS tube,
and the assembly was then baked under 50 mT Argon buffer gas at
200\textcelsius~ for 2 hours. The oven tube was then sealed with the valve and operated at 400\textcelsius~ for over a
year.

\section*{Appendix II: Control Electronics}

\subsection{Temperature control}
\begin{figure}[h]
\centering
\includegraphics[width=4.5in]{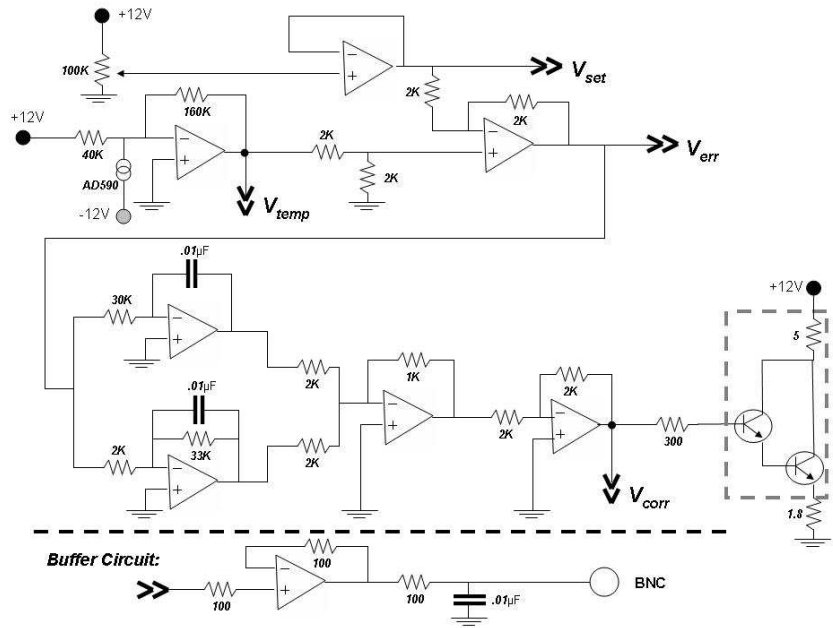}
\caption{Temperature control circuit used to heat laser mount
block. The 5 $\Omega$ resistor and 2 transistors serve as heating elements mounted to the aluminum mounting block. The dashed box indicates which elements are physically mounted onto the aluminum mounting block. The buffer circuit is duplicated four times and utilized where the double-arrows indicate connections to test-points where voltages can be monitored.}\label{temp_circuit}
\end{figure}

The AD590 temperature sensor produces $1 \mu A/ K$. In the circuit
shown in Figure \ref{temp_circuit}, the transimpedance gain and
offset makes $V_{temp}$ equal to 0 V at 20 \textcelsius~ and 10 V at
80 \textcelsius.  The difference between the adjustable $V_{set}$
and $V_{temp}$ makes $V_{err}$. The error signal $V_{err}$ is used
with feedback to regulate $V_{corr}$. The correction signal,
$V_{corr}$ is then used to regulate the current supplied to the heating elements on the laser block (shown within dotted line box in Figure \ref{temp_circuit}). Temperature stabilization circuits with similar function are found in \cite{WIH91,MSW92,RWE95,CAM85}.

\subsection{Peizo gain and offset}
\begin{figure}[h]
\centering
\includegraphics[width=5.5in]{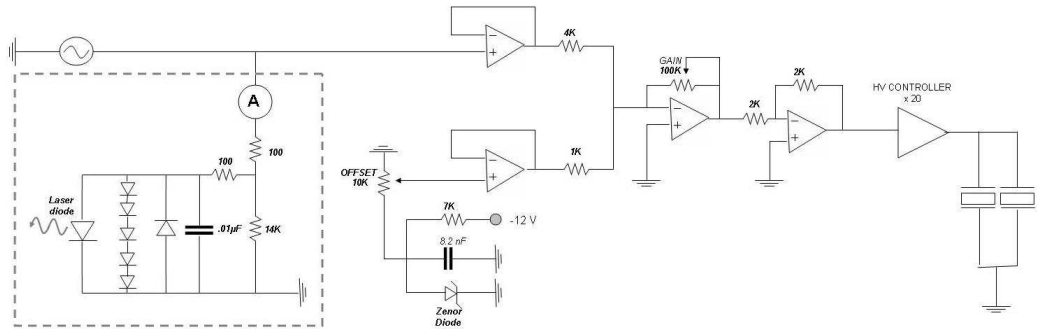}
\caption{Feedforward electronics detail.  The boxed portion of the
circuit shows the laser and its protection circuitry including an ammeter (A).  The potentiometers marked "offset" and "gain"
serve to condition the voltage applied to the piezoelectric elements. }\label{piezo_circuit}
\end{figure}

The piezo elements were controlled with the piezo gain and offset
circuit shown in figure \ref{piezo_circuit}.  This simple circuit
was put in line before the high voltage amplifier (piezoelectric
controller, MDT694A, Thorlabs).

\section*{Appendix III: Initial Alignment Procedure}

These steps are a guide for tuning the cover slip cavity laser frequency to an atomic resonance.
\begin{enumerate}
\item Adjust the laser wavelength to within 0.1 nm of 671 nm by modifying the mounting block temperature while using a grating spectrometer to monitor the wavelength.

\item Ramp the laser current using a 100 Hz triangle function. Make the ramp extend from threshold to the maximum safe current (45 mA in the case of the Sanyo model DL3149-057).

\item Tip and tilt the cover slip by using the kinematic mount to provide optical feedback to the laser cavity.  Do this first by getting the dim secondary laser beam (caused from the back reflection from the cover slip) to overlap with the primary (brighter) laser beam.

\item While observing the transmitted power through a low-finesse Fabry-Perot etalon (the windows of the heat-pipe oven can serve this purpose), verify that the piezo-voltage offset and gain can modify the frequencies at which mode-hops occur.  Re-align the cover slip using control of the laser modes as an indication of good coupling back into the laser cavity.

\item Observe the transmission through the heat-pipe oven with the laser current driven by the 100 Hz ramp.  Slowly ramp the piezo voltage (with the potentiometer) to deliberately hop between different modes, and scan the temperature in order to see any hint of an atomic absorption (or fluorescence) signal.

\item Once any atomic absorption (or fluorescence) has been observed, then obtain a larger mode-hop-free range by following these steps:  (1) adjust the piezo-voltage gain to suppress mode hops, and (2) adjust the piezo-voltage offset to select the laser mode resonant with the atoms.

\item For the largest mode-hop-free tuning range it is often required to iterate steps 1 and 2 and fine tune the laser temperature and the cover slip alignment.

\end{enumerate}

\section*{Acknowledgements}
This work was supported with NSF Grant No. PHY-0653623.  We thank
Dr. W. Bickel and Dr. J. Jones for diagnostic equipment, K. Guerin
for assistance with mechanical drawings, and M. Parker of Rincon
Research Inc. for optics components.


\begin{thebibliography}{10}
\bibitem{laser} The Sanyo DL3149-057, the Mitsubishi ML101-J8 and
the Missubishi ML101-J27 each demonstrated resonant absorption by Li
vapor.  The Sanyo 4039-11 and Hitachi models HL6720, HL6714G, and
HL6724MG were also tested but did not demonstrate resonant
absorption by Li vapor due to unsufficient wavelength tuning range. At least 3 units of each model were tested
over a range of temperature from 0\textcelsius~ to 80\textcelsius. The
Sanyo DL3149-057 gave the largest mode-hop-free tuning range, so
results from this model are discussed here.
\end{thebibliography}
\end{document}